 \documentclass[final,5p,times,twocolumn]{elsarticle}

\usepackage{graphicx}

\usepackage{amsmath}
\usepackage{amssymb}

\journal{Physics Letters B}

\begin{document}

\begin{frontmatter}



\title{First Measurement of the Electromagnetic Form Factor of the Neutral Kaon at a Large Momentum Transfer and the Effect of $SU(3)$ Breaking}


\author[nu]{Kamal~K.~Seth\corref{cor1}}
\ead{kseth@northwestern.edu}

\author[nu]{S.~Dobbs}
\author[nu]{A.~Tomaradze}
\author[nu]{T.~Xiao}

\address[nu]{Northwestern University, Evanston, Illinois 60208, USA}

\author[ws]{G. Bonvicini}

\address[ws]{Wayne State University, Detroit, Michigan 48202, USA}

\linespread{1.5}

\begin{abstract}
At large momentum transfers the photon interacts with the charges and spins of the constituent partons in a hadron.
It is expected that the neutral kaon can acquire finite electromagnetic form factors because its wave function is affected by the order of magnitude difference between the mass of the strange quark and that of the down quark, or flavor $SU(3)$ breaking.
We report on the first measurement of the form factor of neutral kaons at the large timelike momentum transfer of $|Q^2|=17.4$~GeV$^2$ by measuring the cross section for $e^+e^-\to K_SK_L$ at $\sqrt{s}=4.17$~GeV using CLEO-c data with an integrated luminosity of 586~pb$^{-1}$.  We obtain $F_{K_SK_L}(17.4~\textrm{GeV}^2)=5.3\times10^{-3}$, with a 90\% C.L. interval of $(2.9-8.2)\times10^{-3}$.
This is nearly an order of magnitude smaller than $F_{K^+K^-}(17.4~\textrm{GeV}^2)=(44\pm1)\times10^{-3}$, and indicates that the effect of $SU(3)$ breaking is small.  In turn, this makes it unlikely that  the recently observed strong violation of the pQCD prediction, $F_{\pi^+\pi^-}(|Q^2|)/F_{K^+K^-}(|Q^2|)=f_\pi^2/f_K^2$, which is based on the assumption of similar wave functions for the pions and kaons, can be attributed to $SU(3)$ breaking alone.
\end{abstract}


\end{frontmatter}

The quark-gluon structure of hadrons is of the greatest interest for our understanding of Quantum Chromodynamics (QCD) as the fundamental theory of the strong interaction. One of the most important tools for the study of the internal structure of hadrons is the measurement of their electromagnetic form factors at large momentum transfers at which the probe photon sees the charges and spins of the quarks and gluons in the hadron rather than the composite hadron. Measurements of the electromagnetic form factors of the neutral mesons are particularly important because they can acquire finite values only due to their internal structure. Open-flavor neutral pseudoscalar mesons like $K^0$, $D^0$, $B^0$, and $B^0_s$, which contain a quark and an antiquark of different flavors and masses, can have finite electromagnetic form factors. 
Kaons offer one of the best opportunities to study how the different masses of the constituent quarks affect the quark wave functions.  With the strange quark being more than an order of magnitude more massive than the $\langle \textrm{up/down}\rangle$ quarks~\cite{pdg} the $SU(3)$ flavor symmetry is broken in the kaon, and its wave function acquires an antisymmetric component.  According to perturbative QCD (pQCD), measurement of the form factor of the neutral kaon at a large momentum transfer provides a sensitive measure of the effect of $SU(3)$ breaking~\cite{lepage2}.  Unfortunately, the existing measurements of timelike form factors of neutral kaons are limited to small momentum transfers~\cite{novosibirsk-kskl,snd-kskl}.  
These include several measurements made at Novosibirsk by the CMD--2 and SND Collaborations between 1971 and 2006~\cite{novosibirsk-kskl,snd-kskl}.  
All these measurements are limited to $\sqrt{s}<1.38$~GeV, or $|Q^2|<1.90$~GeV$^2$.  
The latest of these~\cite{snd-kskl} reports an average cross section of $\sim0.30$~nb in the region $\sqrt{s}=1.34-1.38$~GeV, which is a thousand times larger than what we report in this paper at $\sqrt{s}=4.17$~GeV.
The ACO and DM1 Collaborations at Orsay have reported measurements at threshold~\cite{novosibirsk-kskl} and in the region $\sqrt{s}=1.400-2.175$~GeV~\cite{novosibirsk-kskl,dm1-kskl}.
For the region, $\sqrt{s}=1.800-2.175$~GeV, $\langle s \rangle\approx4$~GeV$^2$, they report $\langle\sigma\rangle = 0.053\pm0.038$~nb, and $\langle F^2_{K_SK_L} \rangle=0.014\pm0.011$, or $ F_{K_SK_L}(4~\text{GeV}^2)=0.12\pm0.05$.
In this Letter we report on the first measurement of the electromagnetic form factor of the neutral kaon, $F_{K_SK_L}(|Q^2|)$, for the large timelike momentum transfer of $|Q^2|=17.4$~GeV$^2$.
Our measurement provides an estimate of the level of $SU(3)$ breaking in the kaon wave function, and also allows us to comment on the conjecture that $SU(3)$ breaking might be responsible for the recently observed large violation~\cite{ff} of the pQCD prediction that for large momentum transfers the ratio of the form factors of the pion and the kaon, $F(\pi^\pm)/F(K^\pm)$ should be equal the ratio of the squares of their decay constants, $f_\pi^2/f_K^2$~\cite{farrarjackson,lepage1}.

No theoretical calculations of the form factor of the neutral kaon exist. However, Lepage and Brodsky~\cite{lepage2} have pointed out that $SU(3)$ breaking would give rise to an odd asymmetric component in the kaon wave function, and conjectured that if it were large it would lead to a large form factor for the neutral kaon.  Chernyak and Zhitnitsky~\cite{cz} proposed two-humped wave functions for pions and kaons based on QCD sum-rules, and predicted that the antisymmetric component in the kaon wave function due to $SU(3)$ breaking is large. They did not calculate the form factor of the neutral kaon, but found its effect on the form factor of the charged kaon to be quite large.  However, subsequent improved QCD sum-rule calculations predicted that the $SU(3)$ breaking effect on the wave function of the kaon is small~\cite{qcdsum}.  Recent lattice simulations~\cite{braun} and Ads/CFT--based QCD calculations~\cite{brodskyteremond} also find small $SU(3)$ breaking effects.  In view of these developments it is important to determine experimentally the level of $SU(3)$ breaking effect on the form factor of the kaon.

We determine the form factor of the neutral kaon, $F_{K_SK_L}(|Q^2|)$ at $|Q^2|=17.4$~GeV$^2$ by measuring the cross section for $e^+e^-\to K_SK_L$, using data taken with the CLEO-c detector at $\sqrt{s}=4.17$~GeV with integrated luminosity $\mathcal{L}=586$~pb$^{-1}$, which corresponds to 5.54~million $\psi(4160)$ produced.  The detector has been described in detail before~\cite{cleodetector}.  

An important pQCD prediction is that the ratio of the branching fractions for the decay of the vector resonances of charmonium to leptons and hadrons are identical because both are proportional to wave functions at the origin.  This relation was successfully used by us recently to measure the form factors of charged pions and kaons at $\psi(3770)$ and $\psi(4160)$~\cite{ff}.  In the present case, it leads to the following relation between the branching fractions of the $J/\psi$, $\psi(2S)$, and $\psi(4160)$ resonances:
$$ \frac{\mathcal{B}(\psi(4160)\to K_S K_L)}{\mathcal{B}(J/\psi,\psi(2S)\to K_S K_L)} = \frac{\mathcal{B}(\psi(4160)\to\text{electrons})}{\mathcal{B}(J/\psi,\psi(2S)\to\text{electrons})}$$
Using the measured branching fractions for the decays of the  $J/\psi$ resonance, we obtain the estimates $\mathcal{B}(\psi(4160)\to K_SK_L)=5.5\times10^{-8}$ and $N_R(K_SK_L)=0.05$~events, and for $\psi(2S)$ we obtain $\mathcal{B}(\psi(4160)\to K_SK_L)=3.6\times10^{-8}$ and $N_R(K_SK_L)=0.08$~events.  These $\langle N_R \rangle = 0.065\pm0.015$~events due to strong decays of the $\psi(4160)$ resonance constitute a negligible background for the form factor decays.  

\begin{figure}
\begin{center}
\includegraphics[width=3.in]{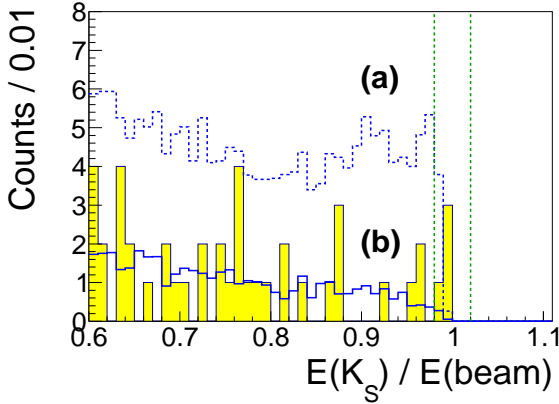}
\end{center}

\caption{JETSET-based Monte Carlo predictions for $e^+e^-\to K_SK_L$ event distributions as function of $X\equiv E(K_S)/E(\text{beam})$: (a) including events in which showers due to $K_L$ energy deposits in the calorimeter are included, (b) events in which no showers are allowed.  Both MC distributions have been normalized to the total number of events in the no shower case in the region $X=0.6-0.98$ to be equal to that in the data distribution, shown as the shaded histogram.}
\end{figure}

\begin{figure*}
\caption{Distributions of the variable $X\equiv E(K_S)/E(\textrm{beam})$ for data taken at (a) $\sqrt{s}=4.17$~GeV, and (b) for comparison, data taken at $\sqrt{s}=3.686$~GeV.   The observed distributions are given by the shaded histograms.  The vertical dashed lines mark the signal region $X=0.98-1.02$. The Monte Carlo determined resolution shapes are shown by the solid line histograms. For (a), $\sqrt{s}=4.17$ GeV, the MC curve has been normalized arbitrarily.}

\begin{center}
\includegraphics[width=3.in]{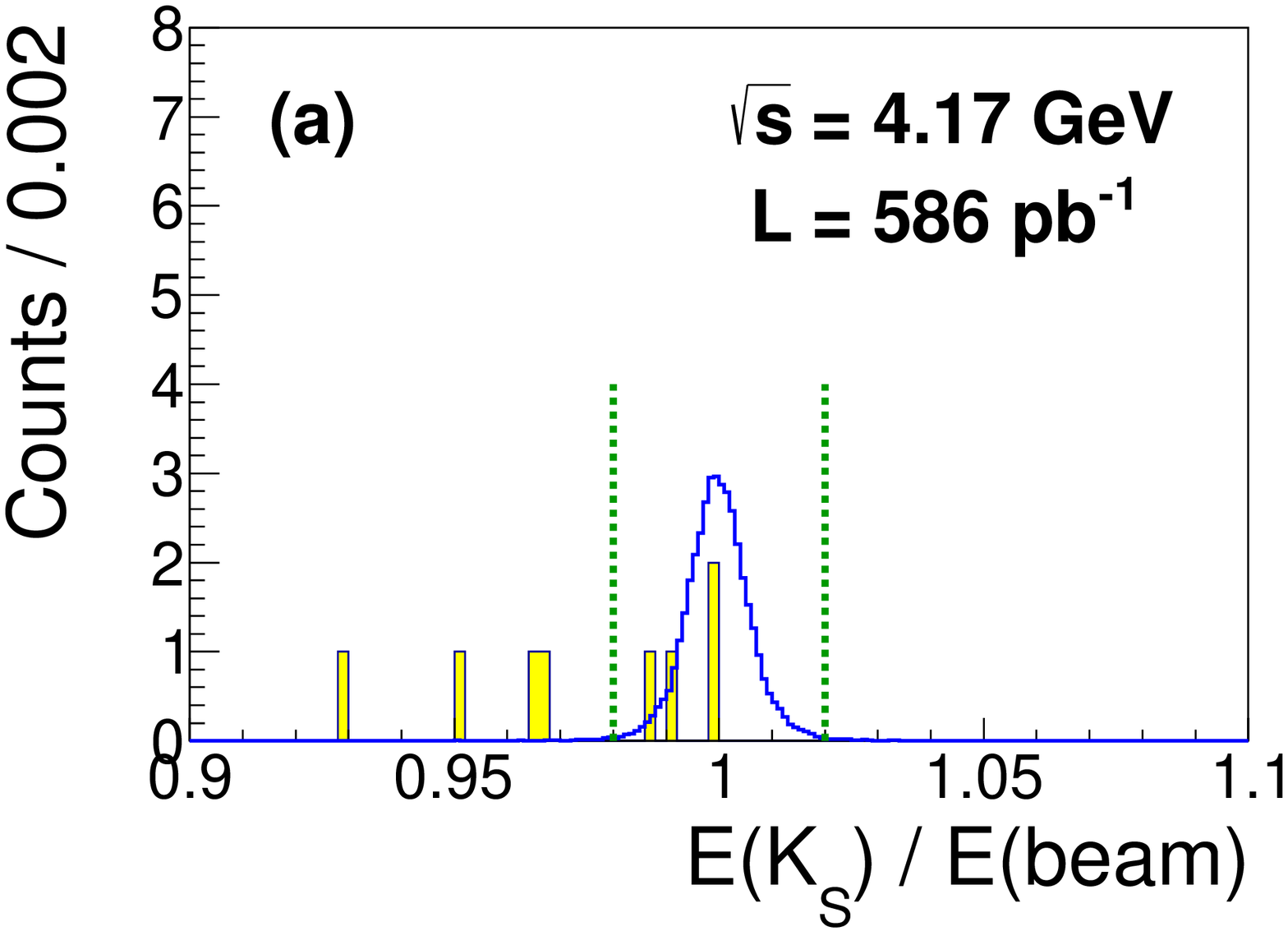}
\includegraphics[width=3.in]{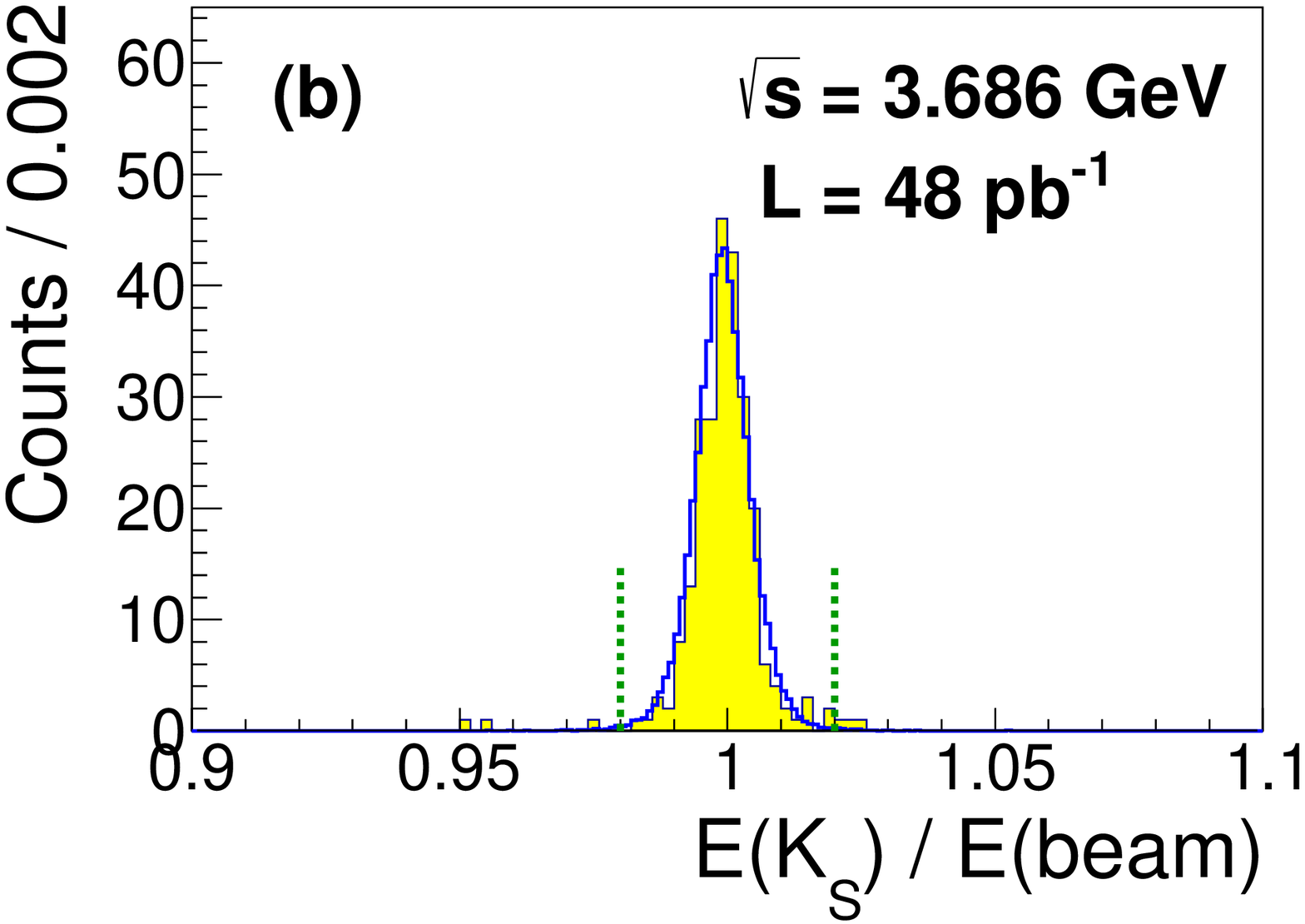}
\end{center}

\end{figure*}

The event selection for $e^+e^-\to K_SK_L$, $K_S\to\pi^+\pi^-$, $K_L(\textrm{undetected})$ is as follows.  Only two charged particles with $|\cos\theta|<0.8$ and zero net charge are allowed in the event, and they are required to meet the standard criteria for track quality.   We identify $K_S$ by its decay into $\pi^+\pi^-$, which can be distinguished from all other charged particles by requiring that the two detected charged particles have their vertex displaced by more than 10~mm from the $e^+e^-$ intersection point, and that their reconstructed total momentum vector extrapolate to within $3\sigma$ of the interaction point.  To further identify the charged pions, we require that their $dE/dx$ in the drift chamber be consistent with the pion hypothesis within $3\sigma$.  To reject electrons, we require that $E(CC)/p<0.9$, where $E(CC)$ is the energy deposited by the charged particle in the crystal calorimeter, and $p$ is the track momentum.

We have examined two alternate procedures for taking account of the fact that the $K_L$ do not stop in the detector, but a fraction ($\sim40\%$) of $K_L$ deposit some of their energy in the central calorimeter.  These shower-producing $K_L$'s can be identified by their direction opposite to that of the identified $K_S$, and such events can be included in our sample of $K_SK_L$. This procedure, detailed in Ref.~\cite{interf}, was used by us to identify the resonance decay, $\psi(2S)\to K_S K_L$, which has a strong signal yield. The other procedure, which is more suitable when the signal yield is small, is to forego all such events by not allowing any showers in the events, and only use events identified by the $K_S$.

We examine the relative merits of the two procedures by Monte Carlo simulations.  For background simulation we use 220~million JETSET generated MC $e^+e^-\to q\bar{q} \to K_SK_L +X$ events.  The normalized distributions of these MC background events are shown in Fig.~1. We note that if the $K_L$ shower identified events are included (histogram marked (a)) the background in the $K_SK_L$ distribution is nearly 4 times larger than when no showers are allowed in the events (histogram marked (b)).  On the other hand, the resulting loss of signal events due to reduced overall efficiency when no showers are allowed in the events is only a factor 1.6, from efficiency, $\epsilon=39.8\%$ to $\epsilon=25.6\%$.  With the very small number of form factor events expected, reduction of the $q\bar{q}$ background is the most important criterion, and we opt for not allowing any showers in the events.  We reconstruct the total energy $E(K_S)$ of the detected $\pi^+\pi^-$, and use the variable $X\equiv E(K_S)/E(\textrm{beam})$ to identify $K_SK_L$ events.  In this variable, the $e^+e^-\to K_SK_L$ events are expected to peak at $X=1.00$.


In Fig.~2 we show two $E(K_S)/E(\textrm{beam})$ distributions observed with the event selection described above.
In Fig.~2(a) we show the distribution for $\sqrt{s}=2E(\textrm{beam})=4.17$~GeV.  In Fig.~2(b), for comparison we also show the distribution for $2E(\textrm{beam})=M(\psi(2S))=3.686$~GeV for CLEO-c data for 24.5~million~$\psi(2S)$.  The $X\equiv E(K_S)/E(\textrm{beam})$ distribution for $\psi(2S)$ illustrates the very clean identification of $K_S$ from $e^+e^-\to K_SK_L$ that our event selection produces.  With this event selection we obtain the branching fraction $\mathcal{B}(\psi(2S)\to K_SK_L)=(5.41\pm 0.36(\text{stat}))\times10^{-5}$, which is in good agreement with $\mathcal{B}(\psi(2S)\to K_SK_L)=(5.28\pm 0.25(\mathrm{stat})\pm0.34(\mathrm{syst}))\times10^{-5}$ reported in our previous measurement in which $K_L$ identified events were included~\cite{interf}.

In the $X \equiv E(K_S)/E(\textrm{beam})$ distribution in Fig.~2(a) at $\sqrt{s}=4.17$~GeV, $s=|Q^2|=17.4~\mathrm{GeV^2}$ we observe 4 events in the signal region, $X=0.98-1.02$.  We also show the arbitrarily normalized Monte Carlo determined peak shape expected for the form factor signal at $E(K_S)/E(\textrm{beam})=1.00$.  
Four scattered events are observed in the off-peak region, $X=0.90-0.98$, but they are insufficient for obtaining a reliable estimate of the background.  
A much more reliable estimate of background can however be made by normalizing the JETSET Monte Carlo distribution for the background to the data in the $X$-region in which the data have a significant number of events. Fig.~1 shows such a distribution normalized with the total number of events in the region $X=0.60-0.98$. It leads to the estimate of 0.3 background events in the signal region, $X=0.98-1.02$. The central value of signal events is thus 3.7. Using the method of Feldman and Cousins~\cite{feldmancousins}, the 90\% confidence interval is 1.3$-$8.3 events, corresponding to 4 observed events with 0.3 background events.

Signal Monte Carlo simulation leads to the determination of event selection efficiency, $\epsilon=25.6\%$. 
The validity of the Monte Carlo simulation for determination of the present reconstruction efficiency is confirmed for $\psi(2S)\to K_SK_L$ (Fig.~2(b)), for which $241\pm16$ events were detected, leading to a branching fraction in agreement with Ref.~\cite{interf}, as stated earlier.
The correction factor for initial state radiation is determined to be $C=0.781$, using the method of Bonneau and Martin~\cite{bonneaumartin}.  The Born cross section  is obtained as 
\begin{equation}
\sigma_B(K_SK_L)=N/[\epsilon\mathcal{L}\:C]
\end{equation}
and the form factor is related to it as,
\begin{equation}
\sigma_B(s,K_SK_L)=(\pi\alpha^2\beta^3/3s)\times|F_{K_SK_L}(s)|^2
\end{equation}
where $\alpha$ is the fine-structure constant, and $\beta = 0.971$ is the velocity of $K_S$ in the laboratory system.
Equations (1) and (2) lead to:
\begin{multline}
\sigma_{B}(17.4~\mathrm{GeV^2},K_SK_L)=0.032~\mathrm{pb},\text{~and} \\
\text{~90\%~C.L.~interval}~0.011-0.071~\text{pb,~and}
\end{multline}
\begin{multline}
F_{K_SK_L}(17.4~\mathrm{GeV^2})=5.3\times10^{-3},~\text{and}\\
\text{~90\%~C.L.~interval}~(3.1-7.9)\times10^{-3}
\end{multline}
These are the first measurements of the form factor and cross section of the neutral kaon at the large momentum transfer, $|Q^2|=17.4~\mathrm{GeV^2}$

As is well known, QCD predicts the dimensional counting rule, according to which the meson form factors decrease as $1/|Q^2|$.  It is interesting to note that the $1/|Q^2|$ extrapolation of the DM1 measurement of $F_{K_SK_L}(4~\text{GeV}^2)=0.12\pm0.05$ leads to $F_{K_SK_L}(17.4~\text{GeV}^2)=(27\pm12)\times10^{-3}$, which is $5\pm2$ times larger than what we measure (Eq.~4).

Our upper limit of $\sigma_B<0.071$~pb is an order of magnitude smaller than $\sigma_B<0.74$~pb reported at $\sqrt{s}=3.67$~GeV~\cite{interf-old}. An earlier CLEO measurement at $\sqrt{s}=3.77$~GeV was compromised by unresolved $K_SK_L$ yield from $\psi(2S)$ populated by ISR, and 8 events were observed in the signal region with a background estimate of 9 events. The 90\% confidence upper limit was quoted as $\sigma_B<0.06$~pb~\cite{kskl3770}.


The systematic uncertainties in the cross section measurement are summarized in Table~\ref{tbl:syst}.  
Luminosity, trigger, track finding, and pion identification uncertainties are identical to those in Ref.~\cite{interf}.  To test the modelling of $K_L$ showering in the calorimeter, we varied the calorimeter selection criteria used to analyze the $\psi(2S)$ data, and found that the maximum variation of the value of $\mathcal{B}(\psi(2S)\to K_SK_L)$ was 10\%.  The uncertainty in the background is determined to be $\pm2\%$ by varying by $\pm25\%$ the interval in which the MC estimate was normalized to the data. The sum in quadrature of the systematic uncertainties is $\pm11\%$, and is dominated by the uncertainty in the efficiency determination.

In Ref.~\cite{ff} we had obtained 
\begin{gather}
\nonumber \sigma_B(17.4~\mathrm{GeV^2},K^+K^-)=2.23\pm0.09(\text{stat})\pm0.12(\text{syst})~\mathrm{pb}, \\
\nonumber F_{K^+K^-}(17.4~\textrm{GeV}^2) = (44\pm1)\times10^{-3}.
\end{gather}

\begin{table}[!tb]
\caption{Systematic uncertainties for $e^+e^-\to K_SK_L$ cross section.}
\label{tbl:syst}

\begin{center}
\begin{tabular}{lc}
\hline \hline
Source & Uncertainty in \% \\
\hline
Luminosity & $\pm1$ \\
Trigger   & $\pm2$ \\
Track finding  & $\pm2$ \\
Pion identification  & $\pm2$ \\
Shower rejection  & $\pm10$ \\
Background  & $\pm2$ \\
\hline
Total   & $\pm11$ \\
\hline \hline
\end{tabular}
\end{center}

\end{table}

Thus the ratios are
\begin{multline}
\sigma_B(K_SK_L)/\sigma_B(K^+K^-)=0.014,~\text{and} \\
\text{~90\%~C.L.~interval}~0.005-0.032
\end{multline}
\begin{multline}
F_{K_SK_L}(17.4~\textrm{GeV}^2) / F_{K^+K^-}(17.4~\textrm{GeV}^2) =0.12,~\text{and} \\
~\text{90\%~C.L.~interval}~0.07-0.19
\end{multline}
It is not possible to confront these results with theoretical predictions because for such large momentum transfers no predictions exist for timelike form factors $F_{K_SK_L}$, $F_{K^+K^-}$, or their ratio.  
However, Lepage and Brodsky~\cite{lepage2} had pointed out that if the ratio of form factors $F_{K_SK_L}(|Q^2|)/F_{K^+K^-}(|Q^2|)$ ``\textit{is indeed appreciable (i.e., of order 1), then the odd, asymmetric components} (which arise due to $SU(3)$~breaking)  \textit{play a major role in the structure of kaon wave function}''.  Our measured value of the ratio, $F_{K_SK_L}/F_{K^+K^-}=0.12$ in Eq.~6, and even the 90\%~CL upper limit of 0.19 are  much smaller than being ``\textit{of order 1}''.  
We therefore conclude that the present measurement implies that the $SU(3)$ breaking effect on the quark wave function of the kaon is small.

The present measurement enables us to also address the question of the relative magnitudes of the form factors of the charged kaon and pion. In our recent precision measurement of these form factors it was found that the ratio $F_{\pi^+\pi^-}(|Q^2|)/ F_{K^+K^-}(|Q^2|) = 1.09 \pm 0.04$ for $|Q^2| = 17.4$~GeV$^2$ . This is in strong disagreement, by more than 9 sigma, with the pQCD prediction that for large momentum transfers this ratio should be equal to the ratio of the squares of the pion and kaon decay constants, $f_\pi^2/f_K^2 = 0.70 \pm 0.01$. Since this pQCD prediction is based on the assumption of similar wave functions for pions and kaons, it was conjectured by Lepage and Brodsky~\cite{lepage2} that its violation could perhaps be explained by the effect of $SU(3)$ breaking in the kaon wave function, if it were found to be large. Our present evidence for a small $SU(3)$ breaking effect in the kaon  wave function makes it unlikely that it is responsible for the  observed large violation of the pQCD prediction for $F(\pi^+\pi^-)/F(K^+K^-)$~\cite{lepage2},  and other possible explanations need to be considered.

To summarize, we have made the first measurement of the form factor of the neutral kaon at $|Q^2|=17.4$~GeV$^2$,  $F_{K_SK_L}(17.4~\textrm{GeV}^2)=5.3 \times10^{-3}$, with a 90\% confidence interval of $(2.9-8.2)\times10^{-3}$, including systematic uncertainties.  This leads to the result that the ratio $F_{K_SK_L}(17.4~\textrm{GeV}^2)/F_{K^+K^-}(17.4~\textrm{GeV}^2)=0.12$, with a 90\% confidence interval of $0.07-0.19$.  It implies that the effect of $SU(3)$ breaking on the wave function of the kaon is small.  In turn this implies that $SU(3)$ breaking can not explain the strong experimental violation of the pQCD expectation that $F_{\pi^+\pi^-}/F_{K^+K^-}=f_\pi^2/f_K^2$ at large momentum transfers.  In all likelihood it is the result of the shortcomings of pQCD, and its validity even at momentum transfers as large as 17.4~GeV$^2$.



This investigation was done using CLEO data, and as members of the former CLEO Collaboration we thank it for this privilege. 
This research was supported by the U.S. Department of Energy.


\bibliographystyle{model1a-num-names}

\begin{thebibliography}{00}



\bibitem{pdg} J. Beringer \textit{et al.} (PDG Collaboration), Phys. Rev. D \textbf{86}, 010001 (2012).

\bibitem{lepage2} G. P. Lepage and S. J. Brodsky, Phys. Rev. D \textbf{22}, 2157 (1980).

\bibitem{novosibirsk-kskl} For a compilation of pre--2003 data, see M.R. Whalley, J. Phys. G \textbf{24},  A1--A138 (2003).

\bibitem{snd-kskl} M.~N.~Achasov \textit{et.~al}, J.~Exp.~Theor.~Phys.~\textbf{103}, 720 (2006).

\bibitem{dm1-kskl} F. Man\'e \textit{et al.} (DM1 Collab.), Phys. Lett. B \textbf{99}, 261. (1981).




\bibitem{ff} Kamal K. Seth, S.~Dobbs, Z.~Metreveli, A.~Tomaradze, and T.~Xiao, Phys.~Rev.~Lett. \textbf{110}, 022002 (2013).

\bibitem{farrarjackson} G. R. Farrar and D. R. Jackson, Phys. Rev. Lett. \textbf{43}, 246 (1979).

\bibitem{lepage1} G. P. Lepage and S. J. Brodsky, Phys. Lett. B \textbf{87}, 359 (1979).


\bibitem{cz} V.~L.~Chernyak, A.~R.~Zhitnitsky and I.~R.~Zhitnitsky, Nucl.~Phys.~B~\textbf{204}, 477 (1982); V.~L.~Chernyak and A.~R.~Zhitnitsky, Phys.~Rep.~\textbf{112}, 173 (1984).

\bibitem{qcdsum} For a summary, see P.~Ball, V.~M.~Braun and A.~Lenz, J.~High~Energy~Phys.~\textbf{05}, 004 (2006).

\bibitem{braun} V.~M.~Braun \textit{et al.}, Phys. Rev. D \textbf{74}, 074501 (2006).

\bibitem{brodskyteremond} S.~J.~Brodsky and Guy~F.~de~T\'eramond, \texttt{arXiv:0802.0514v1[hep-ph]}, and priv.~comm.

\bibitem{cleodetector} See, for example, S.~Dobbs \textit{et al.} [CLEO Collaboration], Phys. Rev. D \textbf{76}, 112001 (2007).


\bibitem{interf} Z. Metreveli \textit{et al.}, Phys. Rev. D \textbf{85}, 092007 (2012).

\bibitem{jetset} T. Sjostrand, Computer Physics Commun. 82 (1994) 74.


\bibitem{feldmancousins} G. J. Feldman and R. D. Cousins, Phys. Rev. D \textbf{57}, 3873 (1998).

\bibitem{bonneaumartin}  G. Bonneau and F. Martin, Nucl. Phys. B \textbf{27}, 381
 (1971).

\bibitem{interf-old} S. Dobbs \textit{et al.} (CLEO Collaboration), Phys. Rev. D \textbf{74}, 011105(R) (2006).

\bibitem{kskl3770}  D. Cronin-Hennessy \textit{et al.} (CLEO Collaboration), Phys. Rev. D \textbf{74}, 012005 (2006); Erratum--ibid. D \textbf{75}, 119903 (2007).

\end{thebibliography}



\end{document}